\documentclass[prl,twoside,twocolumn]{revtex4}
\usepackage{graphicx}
\usepackage{amsmath}

\begin{document}
\title{Trapping in Self-Avoiding Walks with Nearest-Neighbor Attraction}
\author{Wyatt Hooper and Alexander R. Klotz }
\affiliation{Department of Physics and Astronomy, California State University, Long Beach}

\begin{abstract}The statistics of self-avoiding random walks have been used to model polymer physics for decades. A self-avoiding walk that grows one step at a time on a lattice will eventually trap itself, which occurs after an average of 71 steps on a square lattice. Here, we consider the effect of nearest-neighbor attractive interactions on the growing self-avoiding walk, and examine the effect that self-attraction has both on the statistics of trapping as well as on chain statistics through the transition between expanded and collapsed walks at the theta point. We find the trapping length increases exponentially with the nearest-neighbor contact energy, but that there is a local minimum in trapping length for weakly self-attractive walks. While it has been controversial whether growing self-avoiding walks have the same asymptotic behaviour as traditional self-avoiding walks, we find that the theta point is not at the same location for growing self-avoiding walks, and that the persistence length converges much more rapidly to a smaller value.

\end{abstract}

\maketitle 

\section{Introduction}

A self avoiding walk (SAW) is a random walk on which no two sites share the same location. In polymer physics, SAWs are used to model excluded-volume repulsion between two segments of a molecule, and the statistics of self-avoiding walks can be used to make measurable predictions about the scaling behavior of polymers and polymer solutions \cite{mckenzie1971shape}. Typically, the statistics of a self-avoiding walk are averaged over every member of the ensemble that consists of every walk of the same length, assuming that each is equally probable. In contrast, a Growing Self-Avoiding Walk (GSAW) on a lattice can be be generated by taking a step in a random direction from on a lattice site, then a subsequent random step to an open neighboring lattice site, continuing to walk randomly to any unoccupied site until no more sites are available \cite{lyklema1986monte}. While every SAW on a specific lattice can in principle be generated as a GSAW, there are two key differences between the GSAW and traditional SAW ensemble, the first being that GSAWs are likely to reach a state where there are no free adjacent sites and the walk becomes ``trapped'' and terminates \cite{hemmer1984average}, and the second that each N-step walk is not equally likely\cite{lyklema1986monte}. It has been debated as to whether the GSAW belongs to the same universality class as the traditional SAW \cite{majid1984kinetic}. 

Trapping is often considered as a bug rather than a feature; various forms of the Rosenbluth method for generating SAW ensembles bias the walk conditions in order to avoid trapping \cite{grassberger1997pruned}. GSAWs have been proposed as models for polymers whose growth timescales are faster than their relaxation timescales, or more generally as models for polymer growth \cite{hemmer1986trapping}. Recent investigations into active nematic gels based on microstubule polymerization \cite{decamp2015orientational} or so-called ``living'' micellar solutions \cite{korobko2016near} serve as physical examples.
As players of the video game \textit{Snake} are acutely aware, a growing self-avoiding walk will eventually become trapped. This was proven by Hemmer and Hemmer \cite{hemmer1986trapping}, and  simulation of 60,000 GSAWs on a square lattice found that the mean distance reached before trapping was approximately 71 steps, with a probability distribution that peaks near 33 steps. A subsequent thesis by Renner \cite{renner1996self} examined trapping in triangular, honeycomb, and simple cubic lattices (among others), and found that the mean length to trapping is approximately 78, 70, and 4000 respectively. The most detailed numerical simulations of GSAW trapping are those of the enthusiast Hugo Pfoertner \cite{pfoertner}, whose simulations are precise enough to identify strong even-odd asymmetry in the trapping probability distribution. 
In the language of polymer physics, a ``good solvent'' is one in which excluded volume interactions between distant parts of the chain cause a polymer to adopt a ``swollen coil'' configuration that is larger than a non-interacting chain of the same length, described by the statistics of a self-avoiding walk. A ``poor solvent'' is one in which effective attractive interactions between distant parts of the chain dominate over excluded volume and the chain forms a compact ``globule'' that is more compact than an equivalent length non-interacting chain. The coil-globule phase transition between the two occurs in what is known as a ``theta solvent,'' in which the attractive and repulsive interactions exactly cancel and the polymer can be described by the statistics of a random walk \cite{pelissetto2005corrections}.. In practice, the theta point can be found either by fixing the interaction strength and varying the temperature, or by fixing the temperature and varying the interaction strength with the reciprocal effect. An analogous transition between a swollen and compact states may be observed in self-avoiding random walks on lattices, where poor-solvent interactions are implemented in terms of a free energy based on the number of nearest-neighbor contacts between non-adjacent sites on each walk \cite{chang1992surface}, Exact enumeration \cite{ishinabe1985examination} Monte Carlo \cite{meirovitch1989collapse} and renoramlization group \cite{seno1988theta} analysis have identified the theta condition for square lattice SAWs, in terms of the critical self-attraction strength and the asymptotic behavior at the transition.
While attractive interactions are known to change the asymptotic behavior of self-avoiding walks, and the growing self-avoiding walk is known to have different behavior from the traditional self-avoiding walk, there has been no examination of the effects of self-attraction on the growing self-avoiding walk. Here, we examine the effect of nearest-neighbor self-attraction on the growing self-avoiding walk in order to ascertain the correlation between trapping statistics and solvent quality and to provide further insight on the differences between the SAW and GSAW ensembles. 

\section{Theory and Simulations}

\begin{figure}
    \centering
    \includegraphics[width=\columnwidth]{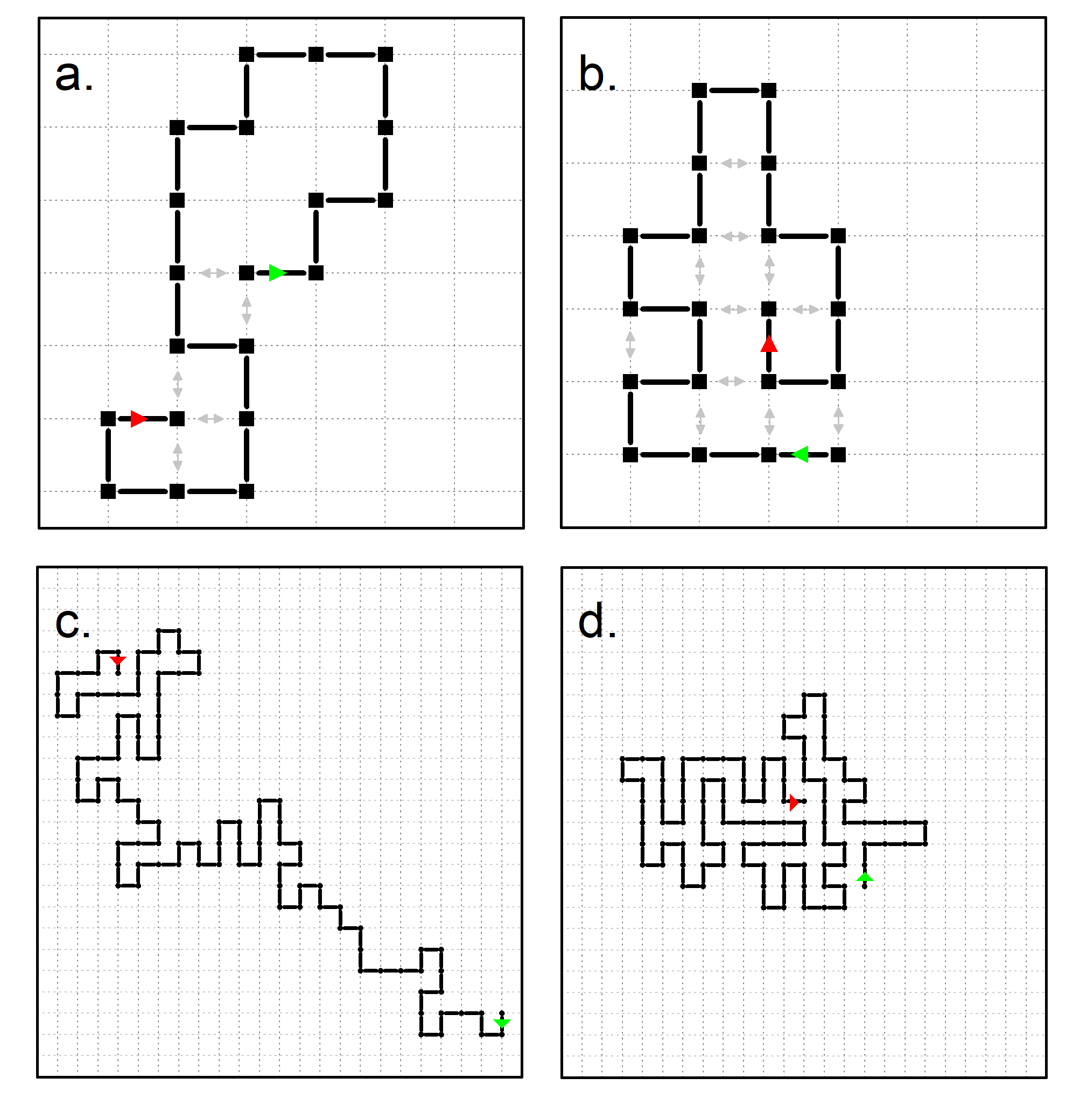}
    \caption{Diagrams of four trapped walks on a square lattice. The top figures are trapped after 19 steps and the bottom after 99. The left figures represent $\beta\epsilon=0$ and the right figures $\beta\epsilon=1$  The first step is indicated with a green arrow, and the final trapped step indicated with a red arrow. Gray arrows indicate the 5 nearest-neighbor contacts in a. and the 10 nearest-neighbor contacts in b. }
    \label{fig:walks}
\end{figure}

A ``walk'' on a square lattice begins at the site $\langle 0,0 \rangle$. It takes its first step towards any of the n=4 nearest-neighbor sites with equal probability. For its second step, there are now n=3 unoccupied adjacent lattice sites. Steps may be taken probabilistically into unoccupied adjacent sites until none remain, at which point the walk is said to have become trapped. The earliest that this can occur is after 7 steps. Figure 1 shows examples of walks that have become trapped after 19 and 99 steps.

For a purely self-avoiding walk, probability of taking a subsequent step to an open lattice site is $p_{o}=\frac{1}{n}$:


More formally, each of the n available steps may be assigned a free energy F such that a partition function Z may be defined:

\begin{equation}
    Z=\sum_{i=1}^{n}e^{-\beta F_{i}},
\end{equation}

the thermodynamic $\beta$ is the reciprocal of the thermal energy, equivalent to 1/T in natural units. The probability of each of the n available steps is:

\begin{equation}
    p_{i}=\frac{1}{Z}e^{-\beta F_{i}}
\end{equation}
 
In a system with no interactions beyond self-avoidance, $F_{i}=0$ and the probability of each step is that of a self avoiding walk, which on a square lattice may be $1/4$ for the first step, and $1/3$, $1/2$, or 1 for each subsequent steps depending on the number of empty adjacent sites. The probability of stepping towards an occupied site is always zero, in contrast to the Domb-Joyce model \cite{barrett1990domb} in which an energetic cost is given to multiply-occupied sites, or the self-avoiding trail model in which only lattice sites but not edges may be shared \cite{lyklema1985growing}.

To study effects analogous to those experienced between non-adjacent monomers in a polymer chain in a poor solvent, we include a negative free energy associated with nearest-neighbor contacts between non-adjacent sites (indicated with gray arrows in Fig. 1 a-b). Such a model is typically used when discussing the theta-point transition in lattice walks. The free energy of each walk in the ensemble depends favorably on the total number $m$ of non-adjacent occupied nearest-neighbor lattice sites, with interaction energy $\epsilon$ per pair:

\begin{equation}
    F_{i}=-\epsilon m.
\end{equation}

The partition function and probabilities may be written:

\begin{equation}
    Z=\sum_{i=1}^{n}e^{\beta\epsilon m_i},
\end{equation}

\begin{equation}
    p_{i}=\frac{1}{Z}e^{\beta\epsilon m_i},
\end{equation}

In simulations parameterized to physical polymers, poor-solvent behavior corresponds roughly to values of $\beta\epsilon$ above 0.5 \cite{pam2007simulation}. In the GSAW ensemble, the number of \textit{new} nearest-neighbor contacts arising from a step will be between 0 and 3. Since each possible state shares a common ``history,'' the number of contacts in the N-1th step may be subtracted from each choice for the partition function of the Nth step. For example, the final step in Fig. 1a gains three new nearest-neighbor contacts. If the last step had been up instead of right, it would have gained 1 new contact, and 0 if it had been to the left. The probability of taking the final step in the right direction would thus be:

\begin{equation}
    p_{right}=\frac{e^{3\beta\epsilon}}{1+e^{\beta\epsilon}+e^{3\beta\epsilon}}
\end{equation}



Growing self-avoiding walks were simulated with MATLAB using parallel CPU computation. The uniform random number generation in MATLAB is based on the Mersenne Twister \cite{moler2004numerical}. Walks were simulated for $\beta\epsilon$ from 0 to 0.6 in increments of 0.01, and up to 3.3 in increments of 0.1. At least 75,000 walks were generated for each value of $\epsilon$, with additional walks generated for specific values of interest. 1.25 million were generated for the $\epsilon$=0 case. For $\beta\epsilon$ values of 6 and 10, 100,000 walks of maximum length 600 were generated to study chain statistics, but it was computationally unfeasible to study trapping at large $\epsilon$.

\section{Results and Discussion: Trapping}

There are several metrics that can characterize the characteristic length of a self-trapping walk, including the mean, median, and peak of the probability distribution. Here, we will primarily discuss the mean trapping length and its dependence on the nearest-neighbor attraction strength.  Other metrics of the trapping distribution are presented in the appendix.

In the simple case of zero self-attraction, we find that the mean trapping length is 70.85 $\pm$ 0.05, consistent with the findings of Hemmer and Hemmer \cite{hemmer1984average}, and within two standard errors of the computations from Pfoertner \cite{pfoertner}. The findings from Renner \cite{renner1996self} are approximately 1.0 greater than Hemmer, Pfoertner, and the present work, suggesting a node-vs-step counting error. 

Trapping requires that the walk first creates a void bounded by occupied sites, and then walks into the void. Voids are unlikely to form in the limit of strong self-attraction, so it may be expected that the mean trapping length increases with the attraction strength. Indeed, we do observe an exponential increase in the mean trapping length with respect to $\epsilon$ (Fig. 1a). For $\beta\epsilon> 1$. the mean trapping length increases exponentially, approximately described by the function $\langle N \rangle\propto e^{1.12\beta\epsilon}$. The largest value of $\beta\epsilon$ for which we measured full trapping statistics was 3.3, at which the mean trapping length was 1152. In 100,000 runs up to length 600, only 120 walks became trapped at $\beta\epsilon=6$ (the lowest at 77 steps) and no trapping events were observed at $\beta\epsilon=10$.

\begin{figure}
    \centering
    \includegraphics[width=\columnwidth]{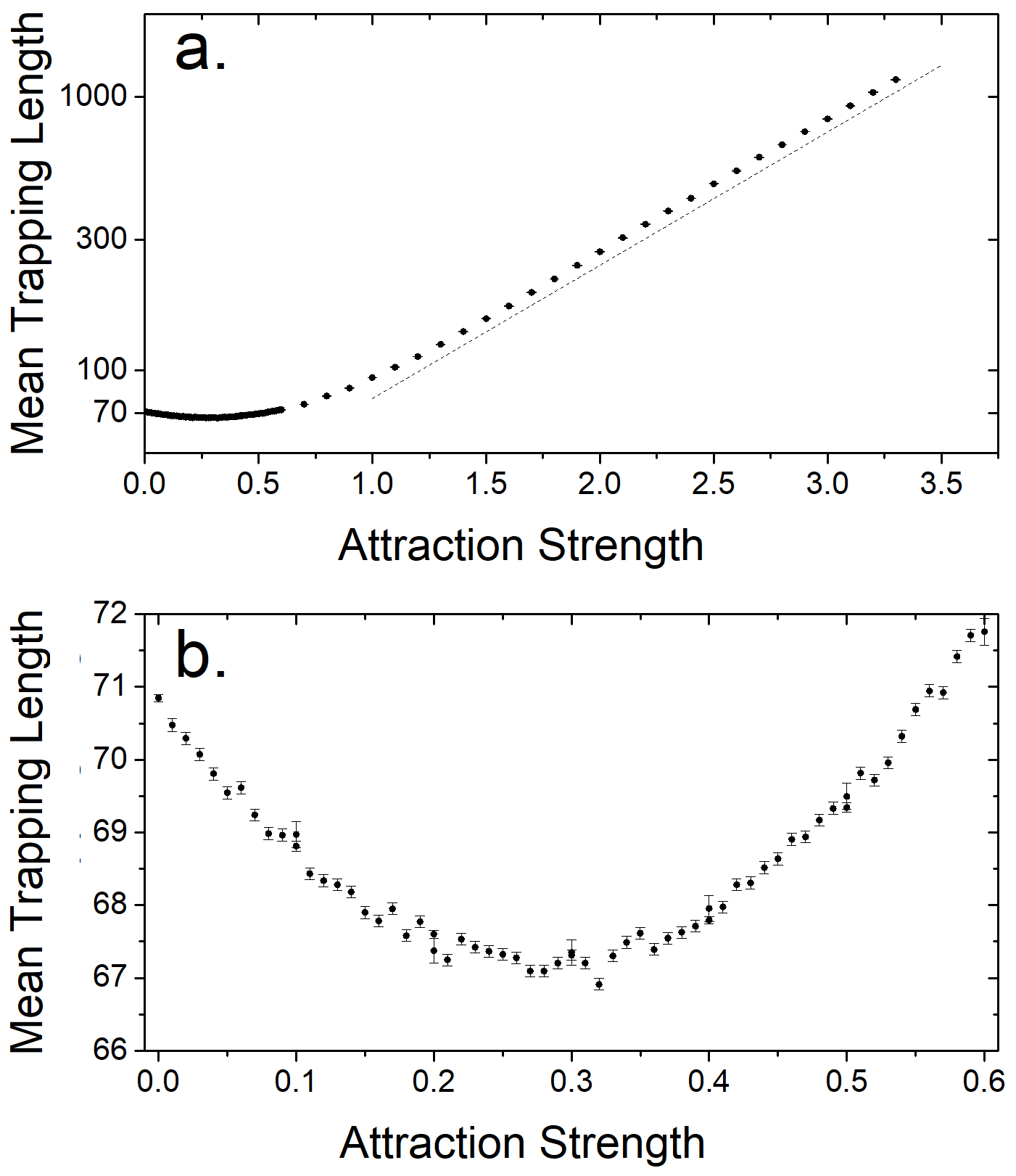}
    \caption{a. Mean trapping length of growing self-avoiding walks as a function of the nearest-neighbor self-attraction strength. Dashed line shows an exponential function with a best-fit coefficient of 1.12. b. The local minimum in trapping length with respect to attraction strength. The minimum is at approximately $\beta\epsilon$=0.28 and $\langle N\rangle=67.2$}
    \label{fig:mean}
\end{figure}

Notably, there is a local minimum at approximately $\epsilon=0.28$ at which the mean trapping length is 67.2 (Fig. 2b). While counterintuitive, the local minimum may be understood by considering that a walk entering a previously formed void will likely have more nearest-neighbor contacts than if it were to avoid the void. When self-attraction is strong, voids are less likely to form for walks to become trapped, but when self-attraction is weak, there is no bias to walk into the voids that do form. The local minimum in the trapping length arises when there is sufficient self-attraction to bias the walks into voids, without suppressing their formation altogether.

\begin{figure}
    \centering
    \includegraphics[width=0.8\columnwidth]{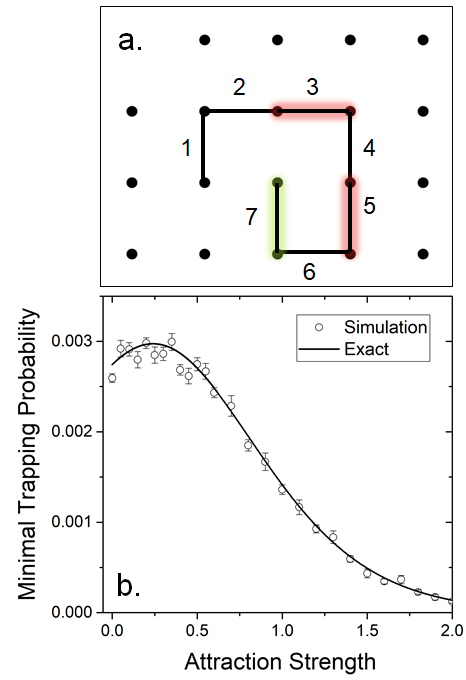}
    \caption{The shortest self-trapping walk lasts 7 steps. a. An example of a minimally-trapped walk. Red lines highlight steps necessary for minimal trapping which are unfavorable because they are not in the direction that increases the number of nearest neighbor contacts. The final green step, highlighted in green, leads to trapping and maximizes the number of nearest-neighbor contacts. b. Calculated probability of trapping after 7 steps, according to Equation \ref{eq:mintrap}, overlaid with simulation results.}
    \label{fig:sev}
\end{figure}

To further understand the local minimum in the mean trapping length, we can consider the the shortest trapped state as an example, which can occur after 7 steps. To become trapped after 7 steps (Fig. 3a), the third and fifth step must both be in an unfavored direction in order to open a void, while the final step must be the most-favored step into the void. The exact probability can of trapping after 7 steps can be calculated from the product of the probability of each necessary step:

\begin{equation}
    P_{7}=\frac{2}{27}\frac{e^{3\beta\epsilon}}{\left(2+e^{\beta\epsilon}\right)\left(2+e^{2\beta\epsilon}\right)\left(1+e^{\beta\epsilon}+e^{3\beta\epsilon}\right)}.
    \label{eq:mintrap}
\end{equation}

This is derived in further detail in the appendix. The minimal trapping probability is non-monotonic with respect to the attraction strength, with a local maximum at approximately 0.234. The minimal trapping probability is well-described by numerical data (Fig. 3b). The maximum in the N=7 trapping probability is analogous to the minimum in the mean trapping length. In principle, one could calculate exact probabilities for longer trapped states and derive the mean trapping length $7P_{7}+8P_{8}+9P_{9}...$, but this becomes difficult as the number of possibilities increases. The minimal trapping probability for the triangular and honeycomb lattices can also be derived, showing that the same local maximum exists in the triangular lattice, but not the honeycomb.

\begin{figure}
    \centering
    \includegraphics[width=\columnwidth]{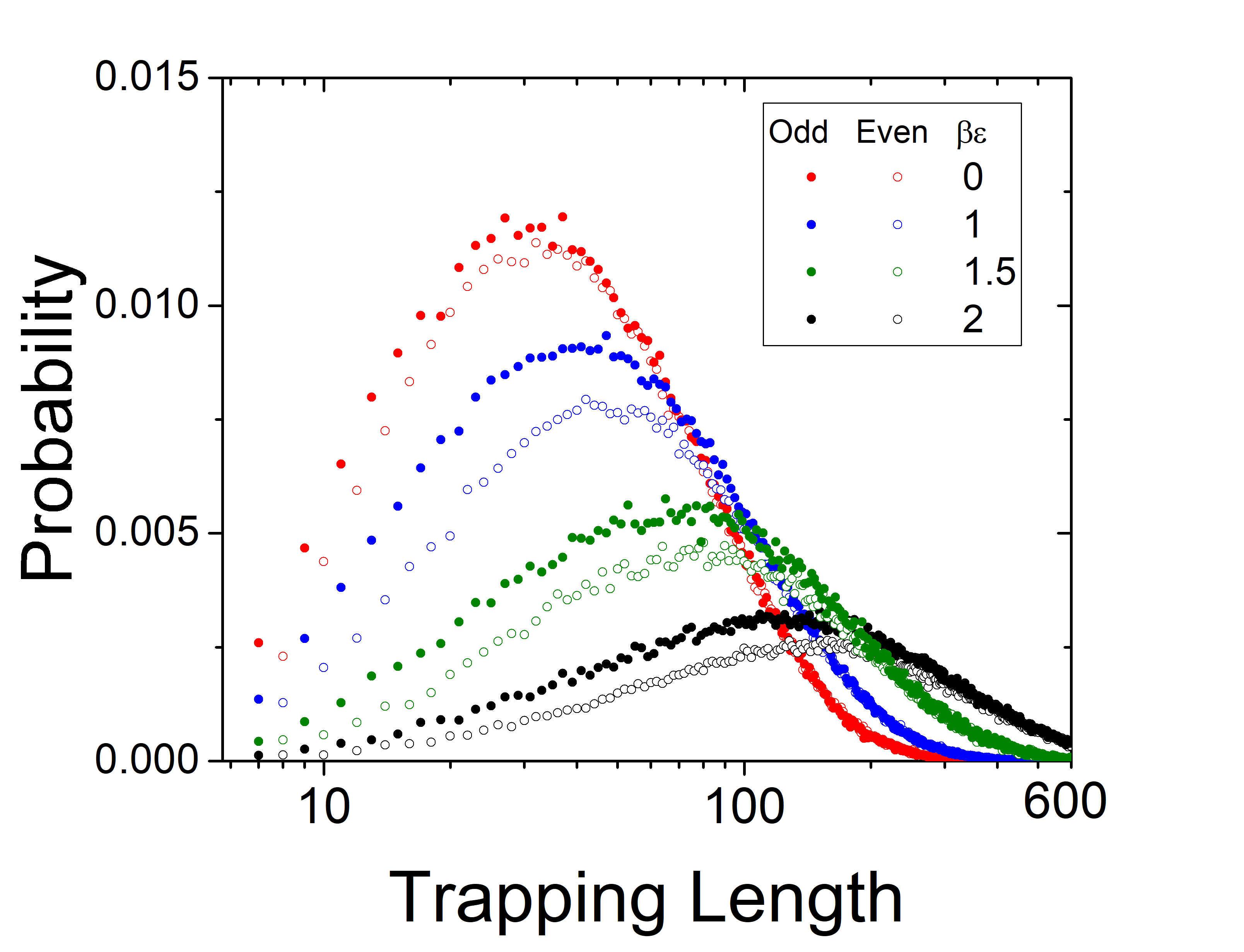}
    \caption{Probability distribution of trapping lengths at $\beta\epsilon$ values of 0, 1, 1.5 and 2. Odd trapping lengths are indicated with solid symbols and even trapping lengths with open symbols. There is strong parity asymmetry which is heightened with increasing self attraction.}
    \label{fig:hists}
\end{figure}

The distribution of trapping lengths was previously described by Hemmer and Hemmer \cite{hemmer1984average} as power-law growth below the peak and exponential decay beyond the peak, finding that for $\epsilon=0$, $p(N)\approx (N-6)^{3/5}e^{-x/40}$. Renner \cite{renner1996self} described similar behavior with a more complex function. We may examine histograms of trapping probability at various $\epsilon$ and see that the peaks are indeed shifted to the right and the tails become broader. We can, in prinicple, make similar fits to histograms and examine how the fit parameters depend on $\epsilon$. This, however, is fundamentally misguided because the trapping probability is not characterized by a single distribution, but two. Even-odd asymmetry in the trapping probabilities has been observed for $\epsilon=0$, with odd-valued trapping being slightly more likely (for example, the probability of trapping after 8 steps is 5/6 that of 7 steps). This can be explained by the abundance of trapped SAWs of various lengths, which itself has even-odd dependence \cite{pfoertner}. The asymmetry becomes more pronounced with increasing $\beta\epsilon$, where histograms can be seen to clearly diverge into to separate curves of different parity (Fig. 4). The probability of trapping in an even state decreases continuously with $\beta\epsilon$, but the distribution of even trapping lengths is typically longer than that of odd: the ratio of the mean of the odd-length to even-length walks reaches a plateau of about 1.05 (Fig. 5). It may be conjectured that the parity asymmetry is an artefact of the square lattice. We also note that the non-monotonic trend in the trapping probability with respect to the attraction strength is not present in the parity asymmetry.

\begin{figure}
    \centering
    \includegraphics[width=\columnwidth]{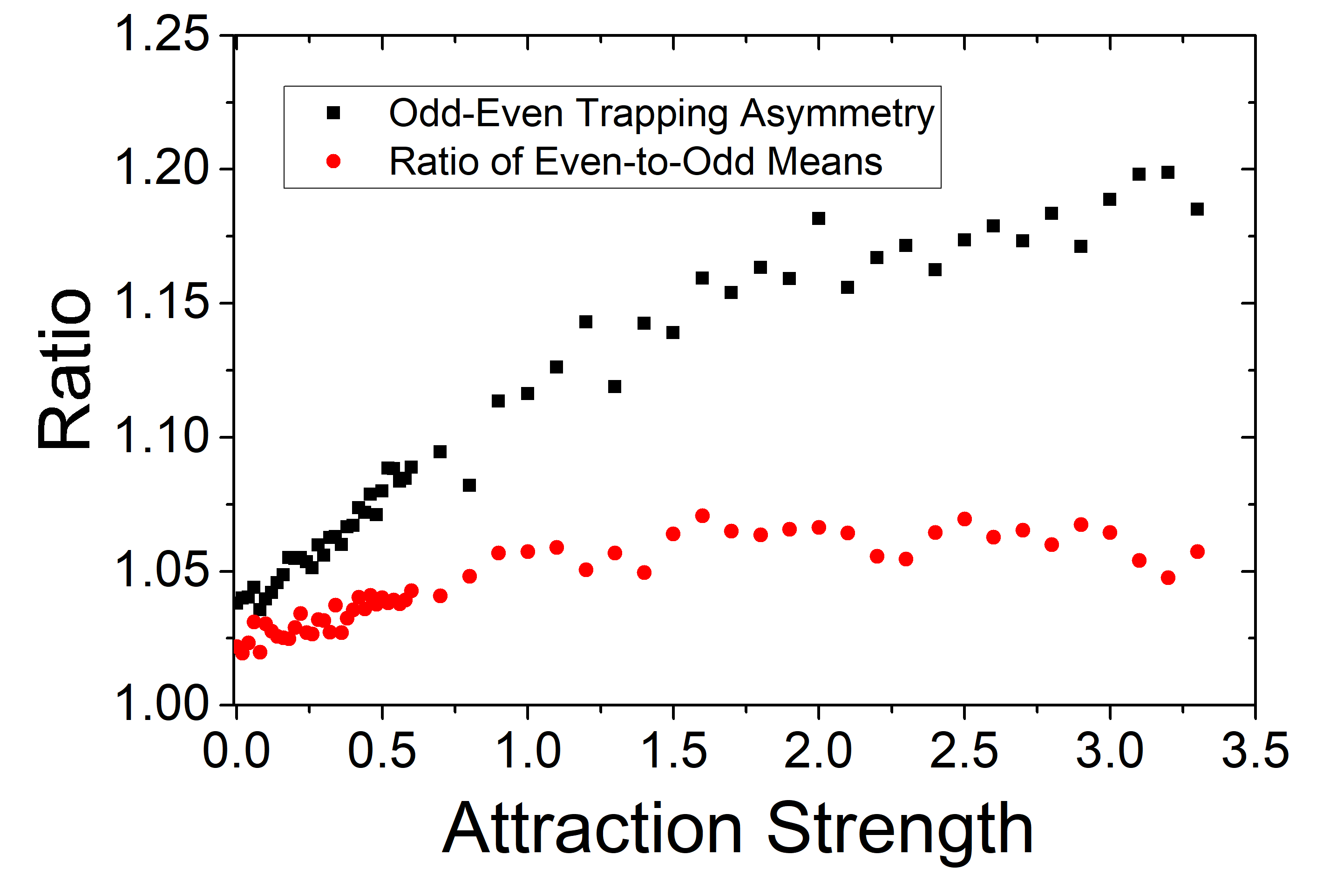}
    \caption{Parity asymmetry as a function of increased attraction strength. Black points indicate the ratio of odd-length trapped walks to even-length trapped walks, which increases continuously. Red points indicate the ratio of the means of the even-length population to the odd-length population, which approaches a plateau near 1.05.}
    \label{fig:parity}
\end{figure}

\section{Results and Discussion: Chain Statistics}

While the primary aim of this study was to examine the effect of self-attraction on self-trapping, we may also examine the statistics of the growing self-avoiding walk ensemble in comparison to traditional self-avoiding walks. We preface this by noting attrition due to self-trapping leads to significantly reduced samples of large walks, which are most important for evaluating critical behavior.

The metrics we use to discuss chain statistics are the radius of gyration, $R_g$, which is the standard deviation of the position of every node in the walk. It may be compared to the end-to-end distance, $R_{ee}$, which is the Pythagorean distance between the origin and the Nth site. Both the end-to-end distance and the radius of gyration are expected to scale according to a power law with respect to N for sufficiently long walks, described by an exponent $\nu$, referred to as the scaling exponent. 

For traditional SAWs in two dimensions, the ``good solvent'' scaling exponent is 0.75, known as the Flory exponent. For compact chains, because the area of a walk is proportional to its length, the exponent is expected to be 0.5. For asymptotically long chains, the exponent is 0.75 at any temperature above the theta point and 0.5 at any temperature below it, with a discrete transition at which point the exponent is 4/7$\approx$0.57. For finitely long chains, there is a smooth transition from 0.75 through 4/7 towards 0.5 \cite{owczarek1994universal}. 

There has been uncertainty in the literature over whether the GSAW follows the same asymptotic behavior as the traditional SAW. Initially it was determined by Majid et al. \cite{majid1984kinetic} to be 0.66, followed by Lyklema and Kremer \cite{lyklema1984growing} who estimated 0.68, both below the Flory value of 0.75. Subsequently, Lyklema and Kremer later argued that it would approach the Flory value for walks that were of such great lengths that it would likely never be seen via simulation \cite{lyklema1986monte}.

To understand why the GSAW may be asymptotically more compact than the SAW, consider the probability of finding a walk extended along a straight line. For the GSAW, the first step may be taken in any direction, and each subsequent step has a 1/3 probability of remaining straight. The probability of finding a length-N GSAW in a straight line is thus $(1/3)^{N-1}$. For a SAW of length N, there are 4 extended walks, and asymptotically $\sim 2.64^{N}$ total walks \cite{grimmett2019self}, meaning the probability of finding a straight path is $4/2.64^{N}$. It is asymptotically much more likely to find an extended SAW than an extended GSAW, and it can be argued that GSAWs are asymptotically more compact.

While we cannot answer the question of asymptotic scaling in the affirmative or negative, we can examine the effect of self-attraction on GSAW statistics in order to identify the swollen-compact transition at the theta point. For a square lattice SAW with nearest-neighbor attraction in two dimensions, the transition occurs at $\beta\epsilon\approx0.66$ with a scaling exponent of 4/7 \cite{caracciolo2011geometrical}. 

\begin{figure}
    \centering
    \includegraphics[width=\columnwidth]{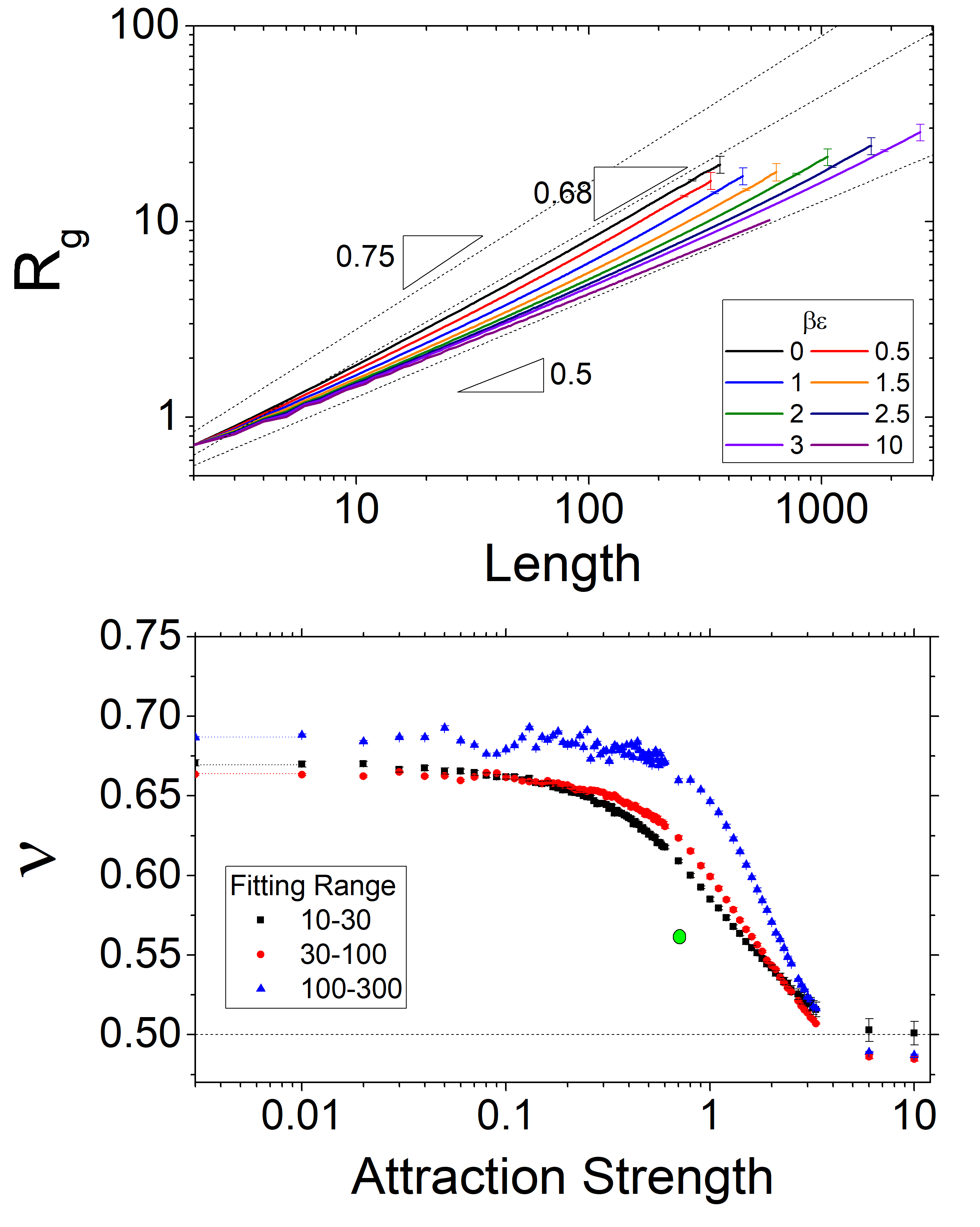}
    \caption{a. Mean radius of gyration of the ensemble on logarithmic axes at $\beta\epsilon$ values of 0, 0.5, 1, 1.5, 2, 2.5 and 3. Data were included until the standard error exceeded 10\%. For clarity, error bars are only shown on the final point. For comparison, power-law curves are shown for the 2D Flory exponent (0.75), the expectation for the collapsed phase (0.5), and the value ascertained by Lyklema and Kremer (0.68) \cite{lyklema1984growing}. b. Best-fit length scaling exponent of $R_g$, plot as a function of the attraction strength. The fits were taken at lengths 10-30, 30-100, and 100-300. The green circle represents the $\theta$ condition for standard self-avoiding walks. $\epsilon=0$ values that would otherwise be absent on a logarithmic axis are shown on the y axis with horizontal lines.}
    \label{fig:rg}
\end{figure}

The scaling of the mean radius of gyration as a function of walk length for several values of $\epsilon$ may be seen in Fig. \ref{fig:rg}a. Generally, the radius scales as a power law, with the exponent decreasing with increasing self attraction. For $\epsilon=0$ the scaling is similar to the 0.68 seen by Lyklema and Kremer \cite{lyklema1984growing}, while for large $\beta\epsilon$ it approaches the 0.5 expected in the compact state. We note that for large $\epsilon$, the behavior of a growing walk is essentially that of an expanding spiral, and many of the traits of an ensemble of ``random'' walks are no longer seen.

Fig. \ref{fig:rg}b shows the results of least-squares power-law fits to various parts of the $R_g$-vs-N data. As expected, there is a smooth transition with $\epsilon$ from between 0.65 and 70 at $\epsilon=0$ to a compact state near 0.5 for large $\beta\epsilon$. The transition is sharper when the fit is made with larger values of N, but does not become discrete. One means of identifying the theta point is to find the value of the attraction strength for which length-dependence vanishes. While the curves appear to converge near $\beta\epsilon=3$, they do not meet at a single point. Rather, the intersection between two $\nu-\epsilon$ curves tends to move towards larger $\epsilon$ with increasing length. We note that the traditional SAW $\theta$ point, at $\beta\epsilon=0.66$ and $\nu=4/7$, is entirely inconsistent with the GSAW data, and we expect from Fig. \ref{fig:rg}b that more data at increased lengths would push the transition towards even larger $\epsilon$. We present additional data to support this in the appendix. Although we have not uniquely identified the $\theta$ point for the GSAW, we can conclusively refute the hypothesis that it is the same as the SAW.

\begin{figure}
    \centering
    \includegraphics[width=\columnwidth]{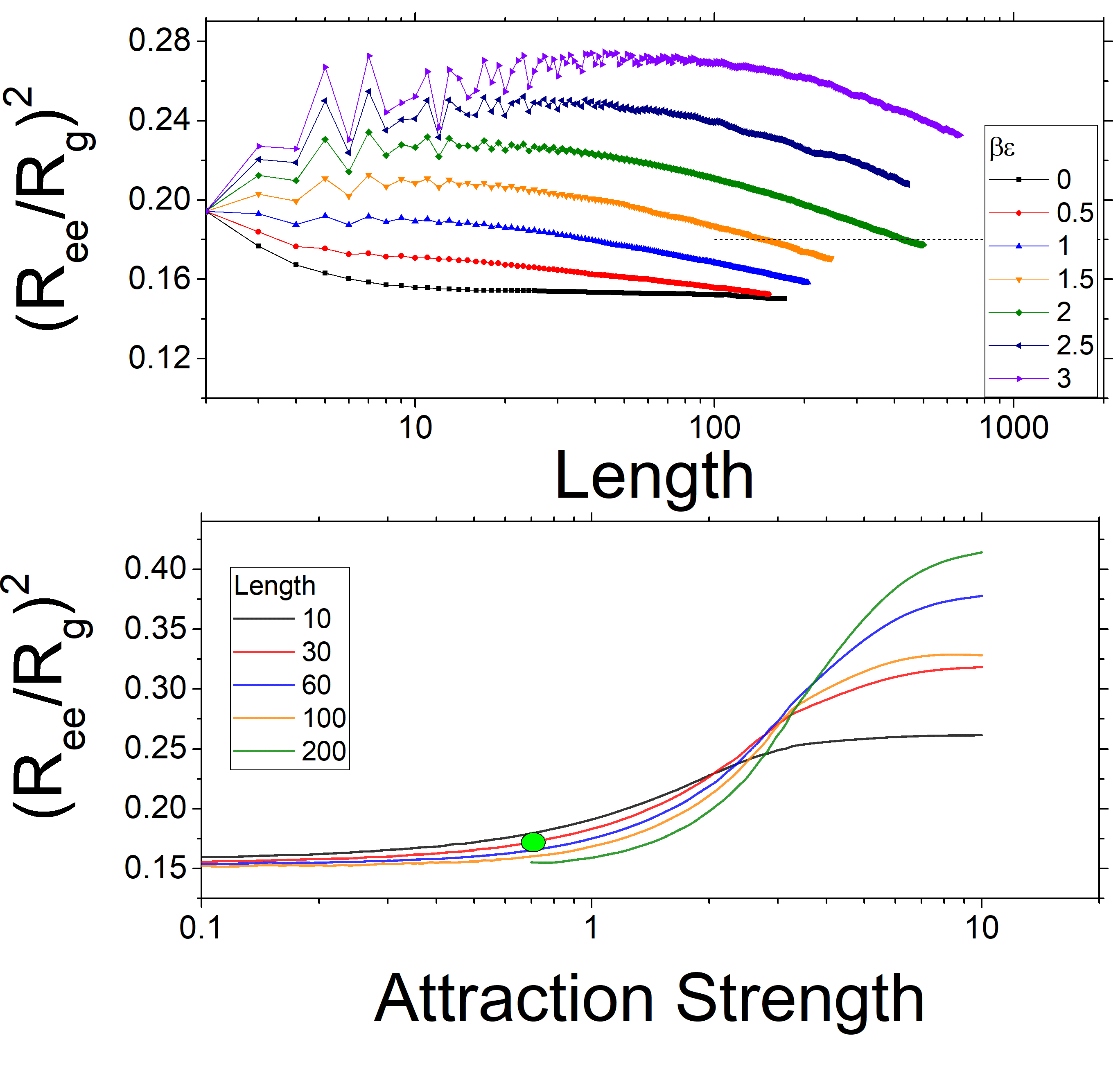}
    \caption{a. Universal amplitude ratio of the squares of the  radius of gyration $R_g$ and end-to-end distance $R_{ee}$ as a function of chain length for several values of attraction strength. The dashed horizontal line represents the value for the SAW theta condition. b. The same ratio plot as a function of the attraction strength for several lengths. The scarce data for $\beta\epsilon$ at N=200 is not shown. The green circle represents the theta point for SAWs.}
    \label{fig:ratio}
\end{figure}

Another metric we can examine is the ratio between the squares of the radius of gyration to the end-to-end distance, also known as a Universal Amplitude Ratio. This ratio reaches a universal constant for asymptotically long polymers, which is approximately 0.18 at the theta point for square lattice walks \cite{caracciolo2011geometrical}. In the swollen regime the ratio is approached from below with respect to length, while in the collapsed regime it is approached from above, and the theta point, there is approximate length independence \cite{grassberger1997pruned}. Examining the data in Fig. \ref{fig:ratio}a it can be seen that the ratio increases with small N for values of $\beta\epsilon$ above 1.5 and decreases below it. For any value of $\beta\epsilon$, however, there is a length beyond which the ratio begins to decrease such that no value of $\epsilon$ will produce a horizontal curve. Fig. \ref{fig:ratio}b shows the ratio at several lengths as a function of the attraction strength with a similar trend to the scaling exponent data, that there is minimal but nonzero length dependence around $\beta\epsilon=3$, but that it moves towards greater $\epsilon$ with increasing length. Similar to the scaling exponent data, there is no single value of $\beta\epsilon$ for which the Universal Amplitude Ratio converges independently of length, and to the extent that there is convergence, it is not at the SAW value of either $\beta\epsilon$ or $R_{g}^{2}/R_{ee}^{2}$.

    
    
    
    

A final parameter of interest is the persistence length of the walk, sometimes known as the persistency to distinguish it from the persistence length of the wormlike chain. The persistence length of a self-avoiding walk represents the ``memory'' a walk has for the direction of its first step. It is defined by Grassberger \cite{grassberger1982persistency} over an ensemble of walks of length N as the mean coordinate of the Nth step in the direction of the first step, or equivalent, the dot product of the position vector of the Nth step and the first step.

\begin{figure}
    \centering
    \includegraphics[width=\columnwidth]{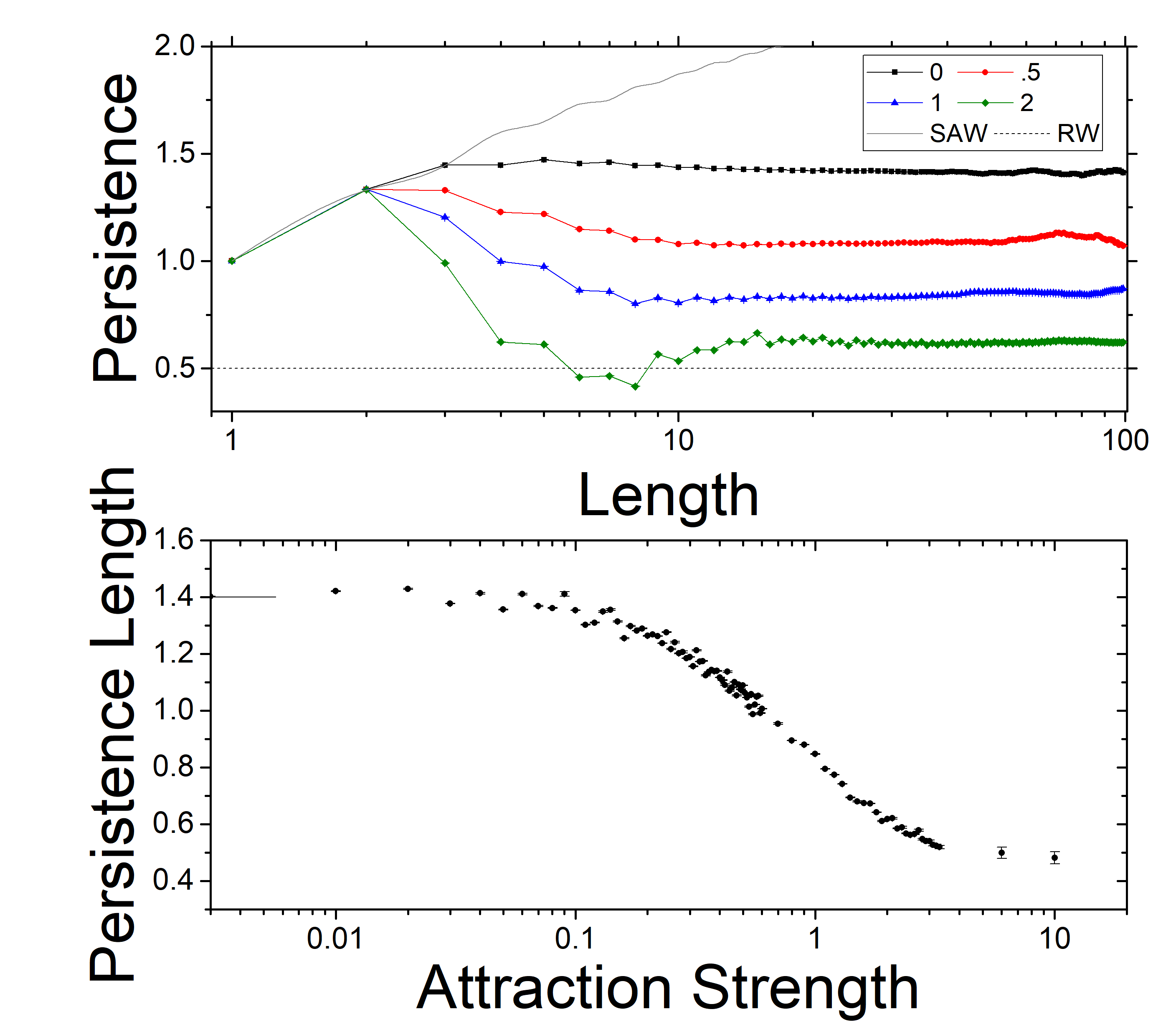}
    \caption{a. The persistence, defined as the mean position of the Nth step in the direction of the first step, as a function of walk length for several values of the attraction strength. The persistence has a convergent value that decreases with attraction strength. For comparison, the random walk value of 0.5 and the values for the traditional self-avoiding walk \cite{granzotti2016scaling} are shown. b. Persistence length as measured by the mean value from length 30 to 60, as a function of attraction strength.}
    \label{fig:per}
\end{figure}

Based on enumerations of self-avoiding walks on the square lattice, Grassberger concluded that the persistence length diverged with a 0.06 power law dependence on length \cite{grassberger1982persistency}. Redner \cite{redner1987persistency} argued using the same data that the persistence was logarithmically divergent, rather than obeying a power law. Eisenberg \cite{eisenberg2003persistence} later used Monte Carlo techniques and examined angular correlation functions to argue that they decayed in such a way as to imply a finite persistence length. Granzotti et al.,  recently extended from Eisenberg's result that the persistence length was in fact 2.66 \cite{granzotti2016scaling}.

Fig. \ref{fig:per}a shows the persistence as a function of length for four values of $\epsilon$. In general, the persistence converges beyond 10 or 20 steps, with noticeable even-odd effects. The convergent value typically decreases with increasing attraction strength. For the $\epsilon=0$ case, the persistence converges to approximately 1.402$\pm$0.001. It is tantalizing but unjustified to ascribe the square root of two to this value. Fig. \ref{fig:per}b shows the persistence length as a function of attraction strength, which decreases smoothly from 1.4 towards approximately 0.5. Compared to the ongoing controversy over the convergence of the SAW persistence, as well as to finite-length effects in other metrics, it is quite clear that the GSAW persistence converges.


\section{Conclusion}

Including nearest-neighbor attraction in the growing self-avoiding walk tends to increase the average number of steps before trapping. However, a local minimum in the trapping length arises for weak values of self-attraction. This can be understand with the idea that a walk must both open a void and step into in order to become trapped, and strongly self-attractive walks do not open voids but weakly self-attractive walks will not necessarily step into them. Investigating the chain statistics of the GSAW ensemble in comparison to the traditional SAW, we have shown that the theta point, if it exists, does not occur at the same value of $\epsilon$, and that the persistence length is convergent and approximately half that of the SAW.
Many of the effects seen in the limit of strong attraction manifest themselves as odd-even effects that may be artefacts of the square lattice and inclusion of only nearest-neighbor interactions. Broadening the analysis to other lattices and including next-nearest-neighbor interactions would be a worthwhile future investigation.
It may be asked whether there is a correlation between trapping statistics and chain statistics, in the same way that the growth of SAW enumerations shares a parameter with its metric scaling \cite{slade2011self}. This is not manifestly the case, as most interesting feature of the trapping statistics occurs at $\beta\epsilon\approx0.25$ and the most interesting feature of chain statistics occurs at $\beta\epsilon\approx3$. We hope that this work encourages more computationally sophisticated investigations into GSAW chain and trapping statistics.

\section*{Acknowledgements}
The authors are grateful to Beatrice W. Soh and Galen Pickett for helpful comments. W.H. is supported by the CSULB Undergraduate Research Opportunity Program.  

\bibliographystyle{unsrt}
\bibliography{walkrefs}

\section*{Appendix}

\renewcommand{\thefigure}{A\arabic{figure}}

\setcounter{figure}{0}

\subsection*{Alternative Trapping Metrics}

Fig A1.a shows the median and distribution peak of the trapping length as a function of the attraction strength alongside the mean. The peak of the distribution is found based on a fit to each trapping histogram using Hemmer and Hemmer's power-exponential function \cite{hemmer1984average}. This measurement is less noisy than the mode of the distribution, but may slightly overpredict the location of the peak (within one step length), in addition to ignoring parity effects. Fig. A1.b shows the ratio of the median and peak to the mean, both of which increase weakly with the attraction strength.

\begin{figure}
    \centering
    \includegraphics[width=\columnwidth]{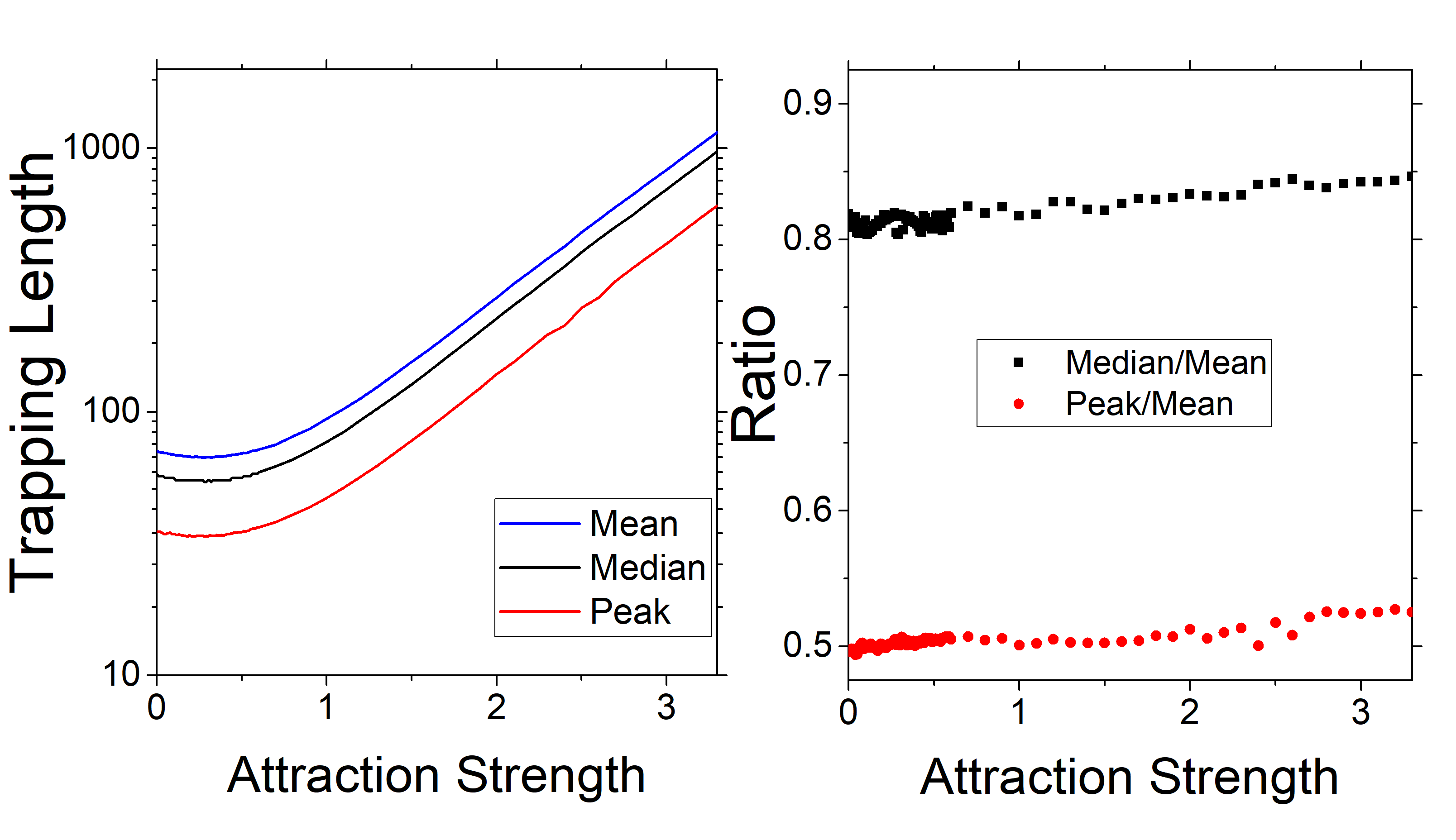}
    \caption{a. Mean, median, and distribution peak of the trapping length as a function of the attraction strength. b. Ratio of the median and the peak to the mean trapping length as a function of the attraction strength.}
    \
\end{figure}

\subsection{Minimal Trapping Probability}

Consider the 7 labelled steps in Fig. 3a. There are eight possible configurations, with four-fold rotational symmetry on the first step and two-fold reflection symmetry on the second step. We will consider the probability that each of the first 7 steps of a walk will be the step required for minimal trapping.
The first step may be taken in any direction.
\begin{align*}
p_1=4\frac{1}{4}.
\end{align*}
The second step may be taken in any of three directions, two of which lead to trapping.
\begin{align*}
p_2=2\frac{1}{3}.
\end{align*}
Assuming without loss of generality that the first two steps are in the +x and +y direction as in Fig. 3a, if the third step is taken in the -y direction it will increase the number of neighbor contacts by 1. The probability of the necessary third step is:
\begin{align*}
p_3=\frac{1}{2+e^{\beta\epsilon}}\leq\frac{1}{3}.
\end{align*}
The fourth step cannot change the number of contacts.
\begin{align*}
p_4=\frac{1}{3}.
\end{align*}
The fifth step is similar to the third in that it must avoid increasing the number of contacts by 2.
\begin{align*}
p_5=\frac{1}{2+e^{2\beta\epsilon}}\leq\frac{1}{3}.
\end{align*}
The sixth step cannot change the number of contacts.
\begin{align*}
p_6=\frac{1}{3}.
\end{align*}
Finally, the seventh step must be in the +y direction to increase the number of contacts by 3 in order to become trapped. An alternative step in the -x direction would increase the number of contacts by 1, and a step in the -y direction would leave it unchanged:
\begin{equation*}
p_7=\frac{e^{3\beta\epsilon}}{1+e^{\beta\epsilon}+e^{3\beta\epsilon}}\geq\frac{1}{3}.
\end{equation*}
The product of each of these probabilities is the total probability of becoming trapped after 7 steps, as outlined in Eq. \ref{eq:mintrap} and plot in Fig. 3b. There are two non-degenerate 8-step trapped walks, the probability of which are left as an exercise to the reader.

    \begin{figure}
        \centering
        \includegraphics[width=\columnwidth]{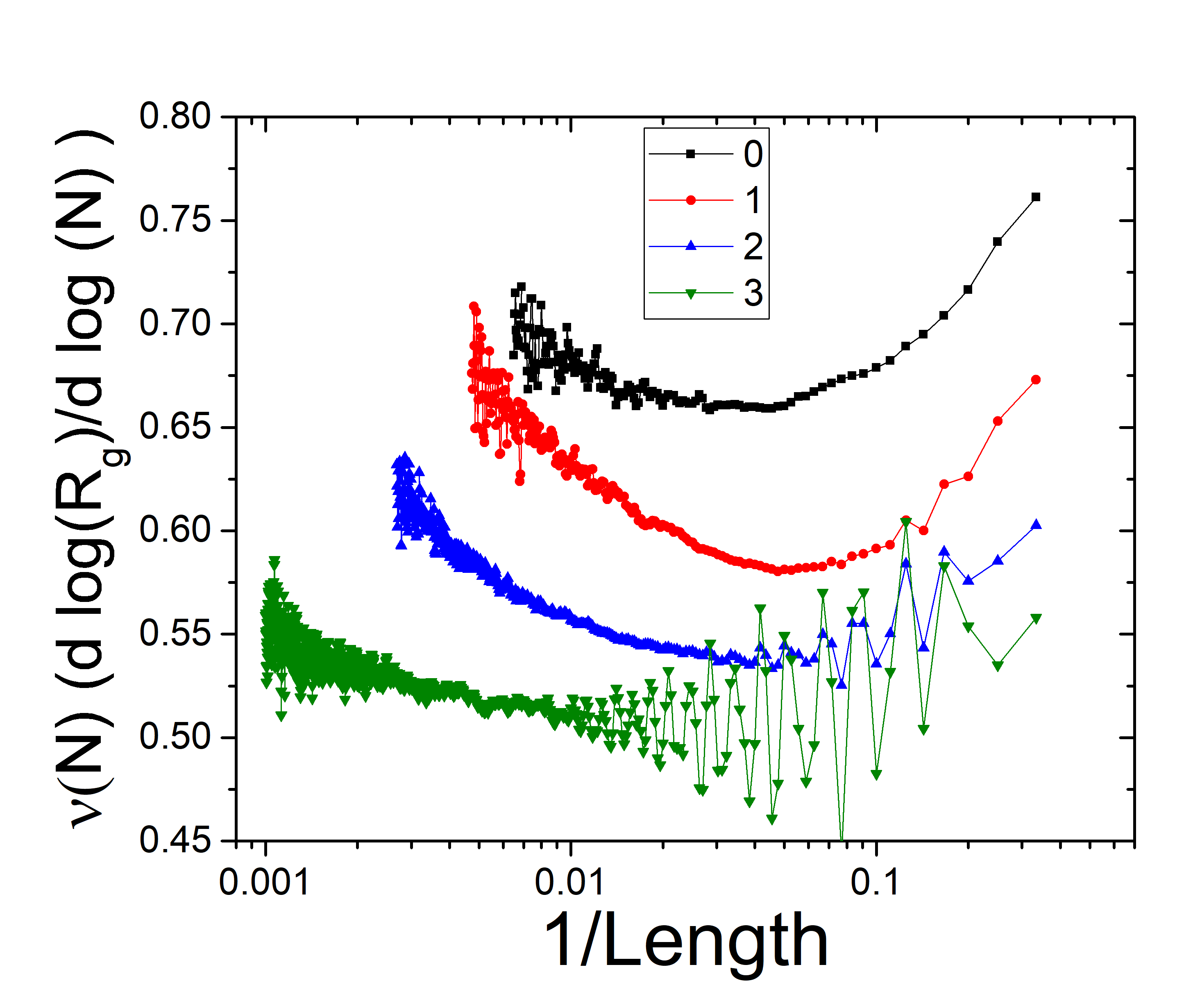}
        \caption{Local scaling exponent of the radius of gyration for several values of $\beta\epsilon$, plot against the reciprocal of the walk length.}
    \end{figure}
\vspace{5mm}
\subsection{Length Dependence of Scaling Exponents}    
Fig. \ref{fig:rg}b shows best-fit scaling exponents at various lengths. Following the conventions of Lyklema and Kremer \cite{lyklema1984growing} and others we can define the ``local'' scaling exponent for walks of a given length as:
\begin{equation*}
    \nu(N)=\frac{\log(R_{g}(N+1)-\log(R_{g}(N-1)}{\log(N+1)-\log(N-1)}.
\end{equation*}
Fig. A2 shows the local scaling exponent at various values of $\beta\epsilon$ as a function of reciprocal length, such that its behavior may be extrapolated towards asymptotically long chains. The data is truncated when it becomes sufficiently noisy. Each curve starts at small N at a local maximum (the ``rodlike'' limit), reaches a global minimum, and then approaches its asymptotic limit as (1/N) approaches 0. The second conclusion of Lyklema and Kremer \cite{lyklema1986monte} is that the asymptotic limit of the $\epsilon=0$ data is 0.75, which we cannot support or refute with our current data. The jagged behavior of the $\beta\epsilon=3$ data is a lattice-induced parity effect. Although each set of data approaches its asymptotic limit from below, if there is a discrete transition at a theta point, we might expect data beyond that point to approach the limit from above \cite{beaton2020two}, which is not observed.


\end{document}